# Towards a Multimodal MRI-Based Foundation Model for Multi-Level Feature Exploration in Segmentation, Molecular Subtyping, and Grading of Glioma

Somayeh Farahani, [1,2,3] Marjaneh Hejazi, [1] Antonio Di Ieva, [3] Emad Fatemizadeh, [4] Sidong Liu [2,3]

1. Department of Medical Physics and Biomedical Engineering, School of Medicine, Tehran University of Medical Sciences, Tehran, Iran.
2. Centre for Health Informatics, Australian Institute of Health Innovation, Macquarie University, Sydney, NSW, Australia.
3. Computational NeuroSurgery (CNS) Lab, Faculty of Medicine, Health and Human Sciences, Macquarie Medical School, Macquarie University, Sydney, NSW, Australia.
4. Department of Electrical Engineering, Sharif University of Technology, Tehran, Iran.

## Abstract

**Background:**
Accurate, noninvasive glioma characterization is crucial for effective clinical management. Traditional methods, dependent on invasive tissue sampling, often fail to capture the tumor's spatial heterogeneity. While deep learning has improved segmentation and molecular profiling, few approaches simultaneously integrate tumor morphology and molecular features. Foundation deep learning models, which learn robust, task-agnostic representations from large-scale datasets, hold great promise but remain underutilized in glioma imaging biomarkers.

**Methods:**
We propose the Multi-Task SWIN-UNETR (MTS-UNET) model, a novel foundation-based framework built on the BrainSegFounder model, pretrained on large-scale neuroimaging data. MTS-UNET simultaneously performs glioma segmentation, histological grading, and molecular subtyping (IDH mutation and 1p/19q co-deletion). It incorporates two key modules: Tumor-Aware Feature Encoding (TAFE) for multi-scale, tumor-focused feature extraction and Cross-Modality Differential (CMD) for highlighting subtle T2–FLAIR mismatch signals associated with IDH mutation. The model was trained and validated on a diverse, multi-center cohort of 2,249 glioma patients from seven public datasets.

**Results:**
MTS-UNET achieved a mean Dice score of 84% ± 3.51 for segmentation, along with AUCs of 90.58% ± 1.25 for IDH mutation, 69.22% ± 3.58 for 1p/19q co-deletion prediction, and 87.54% ± 2.65 for grading, significantly outperforming baseline models ($p \leq 0.05$). Ablation studies validated the essential contributions of the TAFE and CMD modules and demonstrated the robustness of the framework.

**Conclusions:**
The foundation-based MTS-UNET model effectively integrates tumor segmentation with multi-level classification, exhibiting strong generalizability across diverse MRI datasets. This framework shows significant potential for advancing noninvasive, personalized glioma management by improving predictive accuracy and interpretability.

## 1. Introduction:

Gliomas are the most common primary brain tumors in the central nervous system [1]. The 2021 WHO Classification of Tumors of the Central Nervous System now mandates an integrated assessment of histopathological and molecular features—particularly the isocitrate dehydrogenase (IDH) mutation and 1p/19q co-deletion statuses—for accurate diagnosis [2]. Clinical studies indicate that patients with IDH-mutant gliomas, particularly those with low-grade gliomas (LGGs), generally have better outcomes than those with IDH-wildtype tumors [3]. Conventional methods for determining these biomarkers involve invasive procedures like biopsies, which not only carry risks such as bleeding and infection but may also miss the tumor's spatial heterogeneity [4]. This highlights the urgent need for reliable, noninvasive techniques for glioma molecular subtyping and histological grading.

Magnetic resonance imaging (MRI) is the most promising noninvasive modality for glioma evaluation because of its routine clinical use and its ability to capture diverse tissue characteristics through various sequences [5]. For example, IDH-mutant gliomas can have well-defined, nonenhancing margins with small cystic areas on T2-weighted MRI scans and a hyperintense peripheral rim on FLAIR, known as the T2-FLAIR mismatch sign. In contrast, IDH-wildtype gliomas typically show indistinct margins and rim enhancement around central necrosis [6]. Despite these imaging features, accurately predicting molecular subtypes from MRI remains challenging due to the inherent intratumoral heterogeneity—where different regions of the same tumor may exhibit variable genetic and phenotypic profiles [7].

Early approaches to addressing this challenge involved radiomics methods [8,9] which extracted hand-crafted features from manually delineated regions of interest. Although these techniques provided valuable initial insights, their dependence on manual segmentation and feature engineering limited both reproducibility and overall accuracy [10]. More recently, deep learning models, particularly convolutional neural networks (CNNs), have been used to predict glioma subtypes and grades directly from multi-parametric MRI [11], [13]. However, many of these models are single-task systems focusing solely on segmentation or molecular classification, often overlooking tumor-genotype interdependencies [14], as highlighted in a recent systematic review and meta-analysis study [15].

To overcome these limitations, recent research has explored multi-task learning frameworks that jointly perform segmentation and classification [15–18]. For example, Cheng et al. [16] proposed MTTU-Net—a hybrid CNN-Transformer model—that segments tumor regions while predicting IDH mutation status. Nevertheless, balancing the fine-grained details required for segmentation with the global cues necessary for classification led to suboptimal IDH prediction. Similarly, Zhang et al. [19] introduced MFEFnet, which employs a U-Net branch for segmentation and a dedicated encoder with a dual-attention mechanism for IDH classification. Although promising, its focus on a single subtype may restrict its broader clinical application. Collaborative models proposed by Decuyper et al. [17] and van der Voort et al. [18] used U-Net–like architectures to isolate tumor regions and extract shared features for concurrently predicting glioma grade, IDH mutation, and 1p/19q co-deletion status. Yet, these approaches often face challenges related to heavy computational demands and the necessity for large, high-quality datasets to effectively capture both pixel-level and global features.

In response, foundation models have recently emerged as large-scale deep learning architectures pretrained via self-supervised learning to learn robust, task-agnostic representations. By pretraining on large-scale unlabeled datasets, these models address the challenge of limited expert annotations in medical applications, enhancing generalization and ensuring more reliable performance on downstream tasks when labeled data are scarce [20]. Although general medical-specific foundation models such as GatorTronGPT [21], MedSAM [22], and MedCLIP [23] have demonstrated promising performance, specialized models tailored to specific imaging modalities, organs, or diagnostic tasks offer improved accuracy and interpretability while reducing the need for extensive labeled data to refine their learned features for precise diagnostic objectives [20]. However, most existing models have primarily focused on two-dimensional images, such as X-rays, whole-slide images, and fundus images [24]. To our knowledge, only one pioneering study has introduced a foundation model for brain tumor detection and molecular status prediction using MRI [25].

Building on these advancements, we propose a novel multi-task framework that employs a robust, specialized foundation model for comprehensive glioma analysis. Our method is based on the SWIN-UNETR architecture from the BrainSegFounder network, which was originally fine-tuned for brain tumor segmentation [26]. By pretraining with a two-phase strategy on over 42,000 cases of both healthy and diseased brain MRIs, BrainSegFounder achieved superior segmentation accuracy on benchmark datasets. Expanding on this robust model, our Multi-Task SWIN-UNETR (MTS-UNET) framework introduces several key innovations:

1. **End-to-end multi-task learning:** We develop a deep learning model that unifies glioma segmentation, IDH and 1p/19q genotyping, and histological grading within a single framework.

2. **Innovative feature integration**: Our framework incorporates two novel 3D modules: the Tumor-Aware Feature Encoding (TAFE) and the Cross-Modality Differential (CMD). The TAFE module uses segmentation supervision from the SWIN-UNETR's auxiliary branch to guide the extraction of multi-scale, tumor-specific features from the encoder. Meanwhile, the CMD module focuses on extracting and emphasizing subtle imaging cues—specifically the T2-FLAIR mismatch—that are beneficial in distinguishing IDH mutation status.

3. **Robust validation:** We train and validate our approach on a diverse, multi-center dataset comprising 2,249 glioma patients from seven publicly available sources, demonstrating its potential for clinical application.

By combining state-of-the-art foundation deep learning techniques, our multi-task approach aims to overcome the limitations of existing methods and offer a more accurate, generalizable, and clinically applicable solution for glioma segmentation, genotyping and grading.

## 2. Materials and Methods

### 2.1. Patient Population and preprocessing

We retrospectively analyzed preoperative MRI scans from 2,587 glioma patients across seven public datasets: The Cancer Genome Atlas (TCGA), specifically the TCGA-LGG and TCGA-GBM collections [27], Ivy Glioblastoma Atlas Project (Ivy GAP) [28], Río Hortega University Hospital Glioblastoma Dataset (RHUH-GBM) [29], University of Pennsylvania glioblastoma dataset (UPenn-GBM) [30], University of California San Francisco Preoperative Diffuse Glioma MRI (UCSF-PDGM) [31], Erasmus Glioma Database (EGD) [32], and LGG-1p19qDeletion (LGG-1p19q) dataset [33,34].

Patients included both low-grade (grades 2 and 3) and high-grade (grade 4) CNS WHO classifications. Inclusion criteria varied by task. For IDH mutation prediction, patients needed a known IDH status and available T1-weighted (T1), postcontrast T1-weighted (T1C), T2-weighted (T2), and T2-weighted fluid-attenuated inversion recovery (FLAIR) scans. The TCGA and UCSF-PDGM cohorts were used for training and internal validation, while the RHUH-GBM, UPenn-GBM, Ivy GAP, and EGD datasets served as external validation sets. For the 1p/19q co-deletion and tumor grade prediction tasks, inclusion required known 1p/19q status and available T1C and T2 scans, as the LGG-1p/19q dataset included only these modalities. We used the UCSF-PDGM, Ivy GAP, and LGG-1p19q cohorts for training and internal validation of both 1p/19q co-deletion and grading tasks, while the TCGA and EGD datasets were designated as external validation. Additionally, the RHUH-GBM and UPenn-GBM datasets were used only for grade training since they lacked 1p/19q co-deletion labels. For the segmentation task, inclusion criteria required the presence of four preoperative MRI modalities, availability of expert tumor segmentation masks, and no overlap with the BraTS 2021 training set (Figure 1).

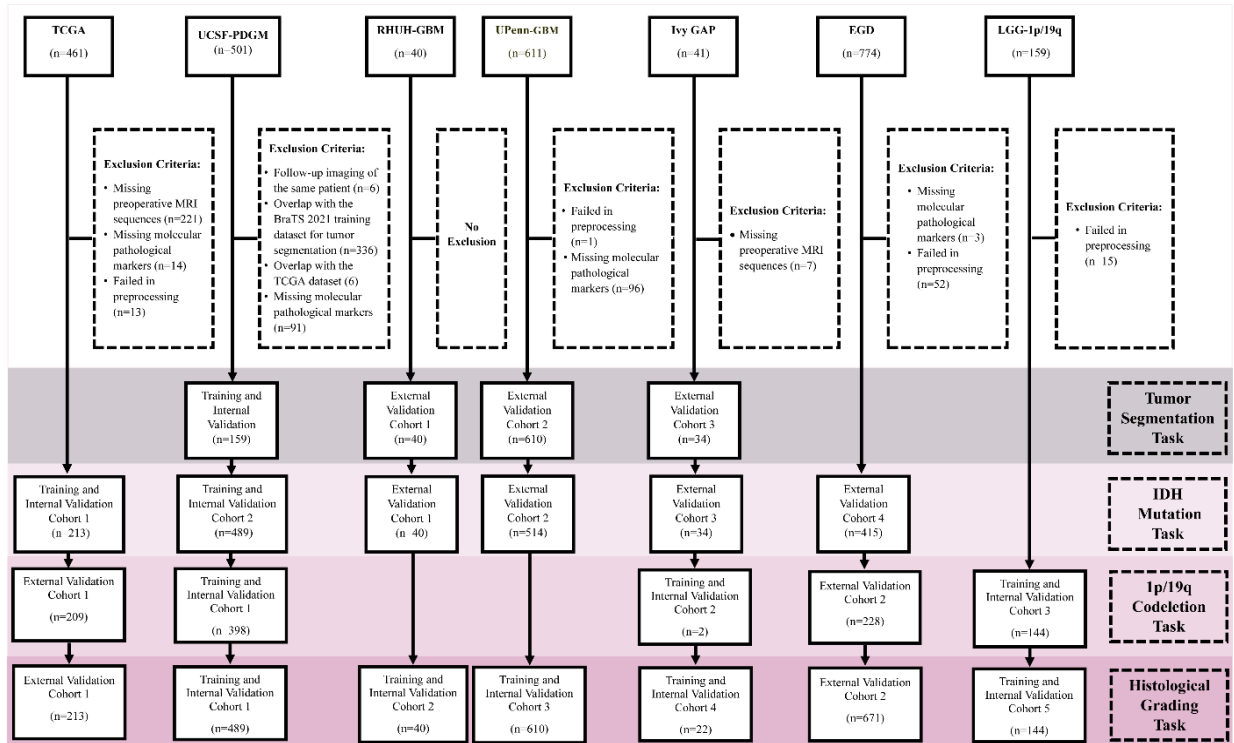

**Figure 1.** Flowchart of patient inclusion and exclusion. The study includes data from multiple sources: TCGA (The Cancer Genome Atlas), Ivy GAP (Ivy Glioblastoma Atlas Project), RHUH-GBM (Río Hortega University Hospital Glioblastoma Dataset), UPenn-GBM (University of Pennsylvania Glioblastoma Dataset), UCSF-PDGM (University of California San Francisco Preoperative Diffuse Glioma MRI), EGD (Erasmus Glioma Database), and LGG-1p19q (LGG-1p19q Deletion). **Abbreviations**: IDH, isocitrate dehydrogenase.

Given the retrospective nature of this study, imaging protocols varied according to local policies. To preserve the inherent clinical heterogeneity, no cases were excluded based on image acquisition parameters or quality. Since the data were acquired from different institutions, various preprocessing methods were applied. For datasets with raw data available (RHUH-GBM, LGG-1p19q, and TCGA), we used the Integrative Imaging Informatics for Cancer Research: Workflow Automation for Neuro-oncology (I3CR-WANO) framework to standardize processing [35]. Whether preprocessed by I3CR-WANO or provided already preprocessed, all datasets underwent a uniform pipeline that included registration to a common anatomical space with a voxel resolution of $1 \times 1 \times 1$ mm³, bias field correction, coregistration to a template atlas, and skull stripping to remove nonbrain tissue. Additionally, all MRI scans were coregistered to T1 or T1C images, normalized using z-score standardization, and cropped to a size of $96 \times 96 \times 96$ voxels.

## 2.2. MTS-UNET framework

Our approach builds on the BrainSegFounder-Tiny model (62M parameters), which employs the SWIN-UNETR backbone—a network that integrates a vision transformer–based encoder with a U-Net–style decoder [26]. Pretrained using a self-supervised dual-phase strategy on large-scale neuroimaging datasets (the UK Biobank [36] with 41,400 healthy brain MRIs and the BraTS 2021 challenge [37] with 1,251 malignant cases), the model was subsequently fine-tuned on the BraTS dataset for brain tumor segmentation. This foundational training enables the model to effectively capture both anatomical and disease-specific features essential for accurate tumor segmentation across various brain regions.

### 2.2.1. Tumor Segmentation

For the segmentation task, we fine-tuned the stage 3 pretrained BrainSegFounder weights [26] on the UCSF-PDGM dataset using a 5-fold cross-validation scheme. To ensure an unbiased evaluation, cases overlapping with the BraTS 2021 training set were excluded. The network produces three nested subregions: GD-enhancing tumor (ET), peritumoral edema (ED), and necrotic/non-enhancing tumor core (NCR/NET). The fine-tuned model was subsequently evaluated on the UPenn-GBM, Ivy GAP, and RHUH-GBM datasets and was also utilized for classification tasks.

### 2.2.2. Molecular and Histological Classification

Figure 2 illustrates our approach, which consists of two modules:

- **TAFE**: Extracts multi-scale, tumor-focused features from all MRI modalities to improve predictions for 1p/19q codeletion, tumor grading, and IDH classification.
- **CMD**: Highlights the subtle T2–FLAIR mismatch sign.

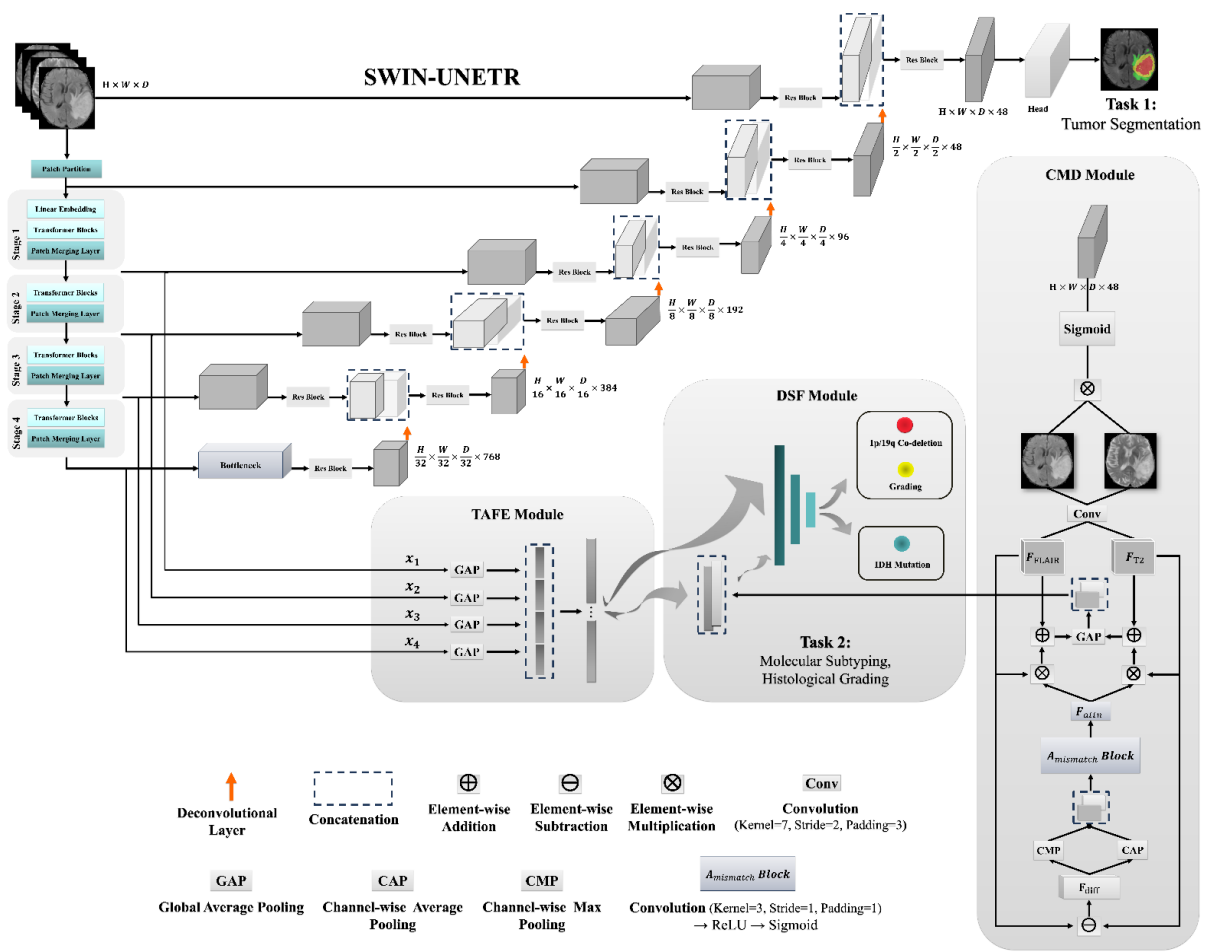

**Figure 2.** Overview of the proposed multi-task learning model. The TAFE module aggregates features from all four stages of the SWIN-UNETR encoder ($x_1$ to $x_4$). This configuration corresponds to one of the four assessed setups of the TAFE module, referred to as TAFE-4, in our ablation study. Numbers in the subgraph represent feature map dimensions.
**Abbreviations**: IDH, isocitrate dehydrogenase.

For IDH prediction, we fuse features from both modules using a Dual-Stream Fusion (DSF) module. For 1p/19q codeletion and grading, only TAFE features are used, as the T2–FLAIR mismatch sign is clinically linked to IDH mutation status. The entire network is trained end-to-end with a joint loss balancing segmentation and classification objectives.

### 2.2.2.1. TAFE Module:

The TAFE module uses tumor segmentation to guide feature extraction across multiple classification tasks (IDH, 1p/19q, and grading). It builds on the SWIN-UNETR architecture as follows:

**Segmentation Branch**

The SWIN-UNETR decoder produces segmentation logits:

$$S \in \mathbb{R}^{B \times 2 \times D \times H \times W},$$

where $B$ is the batch size and $D$, $H$, $W$ denote the 3D volume dimensions. The two channels correspond to background and tumor scores. A segmentation loss (e.g., Dice loss) is applied to $S$ to enforce tumor-specific feature learning.

**Multi-Scale Encoder Feature Extraction**

The SWIN-UNETR encoder generates a hierarchy of feature maps $\{x_1, x_2, x_3, x_4\}$ with

$$x_i \in \mathbb{R}^{B \times d_i \times D_i \times H_i \times W_i},$$

where $d_i$ is the number of channels and $D_i$, $H_i$, and $W_i$ are the spatial dimensions at stage $i$. The segmentation supervision propagates back into the encoder, guiding it to produce representations that are sensitive to tumor regions.

**Feature Fusion for Classification**

Global average pooling (GAP) is applied to selected encoder features:

$$\mathbf{X}_{(i)}^{\text{GAP}} = \text{GAP}(x_i) \in \mathbb{R}^{B \times d_i} \quad (1)$$

and features from one or more stages are concatenated:

$$z = [\mathbf{X}_{(i_1)}^{\text{GAP}}, \mathbf{X}_{(i_2)}^{\text{GAP}}, \ldots] \quad (2)$$

This multi-scale feature vector $z$ is then passed through a fully-connected classification head to predict the relevant tumor attributes, such as 1p/19q co-deletion status, tumor grading, or IDH mutation status.

### 2.2.2.2. CMD Module

The T2-FLAIR mismatch sign is an imaging marker that is readily detectable on standard T2 and FLAIR MRI sequences. Although this sign is highly specific for identifying IDH-mutant gliomas, its sensitivity is limited [38]. To improve detection, we propose a CMD approach that accentuates mismatch patterns in 3D MRI scans. Its operations include:

**Difference Amplification**

Given the T2 and FLAIR volumes T2, FLAIR $\in R^{B \times 1 \times D \times H \times W}$, 3D convolutions yield features $F_{T2}$ = Conv3D (T2) and $F_{FFLAIR}$ = Conv3D (FLAIR). Their difference is amplified as

$$F_{diff} = \gamma \cdot (F_{T2} - F_{FLAIR}) \qquad (3),$$

with $\gamma > 1$ serving as the amplification factor.

**Channel-Wise Pooling and Attention**

To localize regions with pronounced mismatches, we apply channel-wise pooling operations to $F_{diff}$, producing $F_{max}$ and $F_{avg}$. These pooled maps are then concatenated along the channel axis:

$$F_{cat} = [F_{max}, F_{avg}] \in R^{B \times 2 \times D' \times H' \times W'} \qquad (4)$$

To generate an attention map that highlights mismatch-prone regions, we pass $F_{cat}$ through a 3D convolution (kernel size $3 \times 3 \times 3$), followed by a ReLU activation and then a sigmoid function:

$$A_{mismatch} = \sigma (\text{ReLU} (\text{Conv3D} (F_{cat}))) \in R^{B \times 1 \times D' \times H' \times W'} \qquad (5)$$

**Weighted Feature Augmentation**

The computed attention map is then used to enhance the original features. We modulate $F_{T2}$ and $F_{FFLAIR}$ by combining them with the attention map through element-wise multiplication (denoted by $\otimes$):

$$F'_{T2} = F_{T2} + A_{mismatch} \otimes F_{T2} \qquad (6),$$

$$F'_{FLAIR} = F_{FLAIR} + A_{mismatch} \otimes F_{FLAIR} \qquad (7)$$

These augmented features ($F'_{T2}$ and $F'_{FLAIR}$) are further aggregated via adaptive GAP to produce compact, classification-ready vectors.

**Tumor Gating**

To ensure that the enhanced features focus predominantly on tumor regions, the T2 and FLAIR inputs are "gated" using a tumor probability map, $P \in R^{B \times 1 \times D \times H \times W}$, derived from a segmentation network. We define a gating function, $G(P)$, that scales the input while enforcing a lower bound (e.g., min_gate=0.1) to prevent complete suppression:

$$T2_{gated} = T2 \times G(P) \qquad (8),$$

$$FLAIR_{gated} = FLAIR \times G(P) \qquad (9)$$

These gated volumes are then processed through the CMD pipeline.

### 2.2.2.3. DSF Module

While TAFE alone suffices for 1p/19q co-deletion and tumor grading, the accurate prediction of IDH status benefits from complementary information. In the DSF module, we fuse the classification outputs from the TAFE ($C_{TAFE}$) and CMD ($C_{CMD}$) modules:

$$C_{fused} = [C_{TAFE}, C_{CMD}] \qquad (10),$$

and a lightweight multilayer perceptron (MLP) produces the final classification logits:

$$C_{final} = \text{MLP}(C_{fused}) \qquad (11)$$

The network is optimized with a joint loss function:

$$L_{total} = \alpha L_{seg}(S, G) + \beta L_{cla}(C, y) \qquad (12),$$

where $S$ are the segmentation logits, $G$ the ground-truth tumor mask, $C$ the classification logits, $y$ the molecular/histological labels, and $\alpha$, $\beta$ are weighting factors balancing the segmentation and classification tasks.

We employed five-fold cross-validation. In each iteration, four parts were used for training and one for parameter tuning, and this process was repeated five times. The optimal hyperparameters were selected, and the model was then retrained before being tested on independent test sets. Final predictions were calculated by averaging the outputs from the five runs and were reported as mean ± standard deviation. Training was conducted on a single A100 GPU (32 GB RAM) for up to 100 epochs with a batch size of 2, using an Adam optimizer with a learning rate of 1e-4. Early stopping was enabled with a patience of 5 epochs to prevent overfitting and optimize training efficiency. To address data imbalance and enhance generalization, we applied online augmentations such as random flipping, rotation, intensity scaling, and elastic deformation and used a dropout rate of 50% to reduce overfitting.

### 2.3. Evaluation of the Proposed Method and Statistical Analysis

Segmentation performance was measured using the Dice coefficient, Hausdorff distance, and Intersection over Union (IoU). For classification, we assessed the model using accuracy (ACC), sensitivity, specificity, area under the curve (AUC), F1 score, and Matthews correlation coefficient (MCC). The AUC, derived from the ROC curve, measures how well the model distinguishes between positive and negative cases across different thresholds—a higher AUC indicates more stable and threshold-independent performance. The confidence interval was computed using the DeLong method [39], which accounts for the covariance of sensitivity and (1 − specificity). The F1 score balances precision and recall, while MCC is particularly useful for imbalanced datasets as it considers both true positives and true negatives. MCC values range from -1 (inverse prediction) to +1 (perfect prediction), with 0 indicating random performance. Higher values across these metrics indicate better model performance.

For the segmentation task, our method was compared to the pretrained SWIN-UNETR model [40]. For classification tasks, our method was compared to previous glioma prediction approaches, including foundational Vision Transformer (ViT) (with 4-block and 8-block configurations) [25], SENet101 [41], ResNet10 [42], ResNet50 [19,43], and DenseNet121 [44]. For a fair comparison, the ViT model was trained using its self-supervised pre-trained weights, while the other models were trained and tested both with and without ImageNet pre-trained weights. The best-performing results for each model were reported. Since the data were normally distributed (per the Shapiro-Wilk test), we applied ANOVA with post-hoc pairwise comparisons ($p < 0.05$) to confirm our method's superior and robust performance for the AUC metric. All statistical analyses were conducted in R (version 4.4.1) using the rstatix package.

### 3. Results

Figure 1 outlines the patient selection process and the distribution of cohorts for each task. A total of 2,249 patients were included across four tasks: glioma segmentation, IDH mutation prediction, 1p/19q co-deletion detection, and histological grading. Table 1 provides an overview of patient characteristics, the availability of conventional MRI modalities, and expert-segmented tumor labels for the datasets. For the segmentation task, 159 patients were used for cross-validation, while 684 cases were allocated for independent testing. The IDH mutation prediction task included 1,705 cases, divided into five sets for cross-validation and independent testing. The 1p/19q co-deletion task used 981 cases, while histological grading was performed on 2,189 cases.

**Table 1.** Patient characteristics, availability of conventional MRI modalities, and expert tumor segmentation labels across the seven datasets: TCGA (The Cancer Genome Atlas), Ivy GAP (Ivy Glioblastoma Atlas Project), RHUH-GBM (Río Hortega University Hospital Glioblastoma Dataset), UPenn-GBM (University of Pennsylvania Glioblastoma Dataset), UCSF-PDGM (University of California San Francisco Preoperative Diffuse Glioma MRI), EGD (Erasmus Glioma Database), and LGG-1p19q (LGG-1p19q Deletion dataset). Class distributions for molecular and histological grades are reported as counts and percentages. Glioma grades 2 and 3 are categorized as low-grade glioma (LGG), and grade 4 as high-grade glioma (HGG).

| Datasets | TCGA (n = 213) | UCSF-PDGM (n = 489) | EGD (n = 719) | Ivy Gap (n = 34) | UPENN-GBM (n = 610) | RHUH-GBM (n = 40) | LGG-1p/19q (n = 144) |
|---|---|---|---|---|---|---|---|
| **Grade** | | | | | | | |
| 2 | 47 (22%) | 56 (11%) | 119 (16%) | 0 | 0 | 0 | 92 (64%) |
| 3 | 59 (28%) | 43 (9%) | 78 (11%) | 0 | 0 | 0 | 52 (36%) |
| 4 | 107 (50%) | 390 (80%) | 474 (66%) | 22 (65%) | 610 (100%) | 40 (100%) | 0 |
| Unknown | 0 | 0 | 48 (7%) | 12 (35%) | 0 | 0 | 0 |
| **IDH Status** | | | | | | | |
| Mutated | 89 (42%) | 103 (21%) | 139 (20%) | 3 (9%) | 16 (2%) | 4 (10%) | 0 |
| Wildtype | 124 (58%) | 386 (79%) | 276 (38%) | 31 (91%) | 498 (82%) | 36 (90%) | 0 |
| Unknown | 0 | 0 | 304 (42%) | 0 | 96 (16%) | 0 | 144 (100%) |
| **1p/19q Co-deletion Status** | | | | | | | |
| Co-deleted | 27 (13%) | 15 (3%) | 68 (10%) | 2 (6%) | 0 | 0 | 94 (65%) |
| Intact | 182 (85%) | 383 (78%) | 160 (22%) | 0 | 0 | 0 | 50 (35%) |
| Unknown | 4 (2%) | 91 (19%) | 491 (68%) | 32 (94%) | 610 (100%) | 40 (100%) | 0 |
| **MRI Modalities** | T1, T1C, T2, FLAIR | T1, T1C, T2, FLAIR | T1, T1C, T2, FLAIR | T1, T1C, T2, FLAIR | T1, T1C, T2, FLAIR | T1, T1C, T2, FLAIR | T1C, T2 |
| **Expert Segmentation masks** | - | ✓ | ✓ (Only WT region) | ✓ | ✓ | ✓ | ✓ |

**Abbreviations:** IDH, isocitrate dehydrogenase; WT, whole tumor.

## 3.1. Tumor Segmentation

The MTS-UNET model achieved high performance in 5-fold cross-validation on the UCSF-PDGM dataset, with a mean Dice score of 87% ± 1.82. In external validation on the Ivy GAP, UPenn-GBM, and RHUH-GBM datasets, it obtained a mean Dice score of 84% ± 3.51, a Hausdorff distance of 4.13 ± 1.96 mm, and an IoU of 79% ± 2.51 across all three tumor subregions (ET, ED, and NCR/NET) (Table 2). The proposed MTS-UNET network outperformed the SWIN-UNETR model across all metrics, demonstrating statistically significant improvements in the Dice score (for the ED sub-region) and IoU (for the ET sub-region) ($p \leq 0.05$). Figures 3 visualize the ground truth and predicted segmentation results overlaid on the FLAIR scans.

Table 2. Performance evaluation of 5-fold cross-validation on the UCSF-PDGM (University of California San Francisco Preoperative Diffuse Glioma MRI) dataset and external validation on the Ivy GAP (Ivy Glioblastoma Atlas Project), UPenn-GBM (University of Pennsylvania Glioblastoma Dataset), and RHUH-GBM (Río Hortega University Hospital Glioblastoma) datasets. Metrics include mean Dice score, Hausdorff distance, and IoU values. The best-performing mean values are highlighted in **bold**. Asterisks indicate statistical significance compared to MTS-UNET, where *$p < 0.05$.

| **Internal Validation** | | | | | **Dice Score (%)** | | | | |
|---|---|---|---|---|---|---|---|---|---|
| **Model** | | **ET** | **ED** | **NCR/NET** | | | **Avg.** | | |
| **MTS-UNET** | Fold 1 | **91** | **90** | 87 | | | **89** | | |
| | Fold 2 | 89 | **90** | 83 | | | 87 | | |
| | Fold 3 | 89 | 89 | 79 | | | 86 | | |
| | Fold 4 | 90 | 86 | 82 | | | 86 | | |
| | Fold 5 | 87 | 86 | 79 | | | 84 | | |
| | **Avg.** | **89** | 88 | 82 | | | 87 | | |
| **External Validation** | | | | | | | | | |
| | | **Dice Score** (mean±std) (%) | | | **Hausdorff** (mean±std) (mm) | | | **IoU** (mean±std) (%) | | |
| | | **ET** | **ED** | **NCR/NET** | **ET** | **ED** | **NCR/NET** | **ET** | **ED** | **NCR/NET** |
| SWIN-UNETR | | 82 ± 2.08 | 82 ± 3.31* | 78 ± 2.51 | 2.01 ± 2.55 | 4.95 ± 2.02 | 7.01 ± 3.01 | 75 ± 2.46* | 80 ± 2.11 | 74 ± 3.10 |
| **MTS-UNET** | | **84 ± 3.51** | **87 ± 2.11** | **80 ± 3.17** | **1.97 ± 1.01** | **4.61 ± 0.71** | **5.81 ± 1.98** | **79 ± 2.58** | **81 ± 1.87** | **76 ± 3.31** |

**Abbreviations:** ET, enhancing tumor; ED, peritumoral edema; NCR/NET, necrotic/non-enhancing tumor core; IoU, Intersection over Union.

## 3.2. Molecular and Histological Classification

Table 3 presents the mean ± standard deviation of key performance metrics—ACC, F1-score, MCC, and AUC—for three classification tasks: IDH mutation, 1p/19q co-deletion, and tumor grading. Each model was evaluated under both 5-fold cross-validation and external validation. The proposed MTS-UNET demonstrated higher performance across most metrics compared to baseline models, with statistically significant improvements ($p \leq 0.05$). While some models, such as ResNet10, SENet101, and ViT-16, achieved competitive results in cross-validation, their performance declined significantly on external test sets. They specifically struggled to balance MCC and F1 scores, which is crucial given the highly imbalanced class distribution in the test cohorts. Figure 4 visualizes occlusion sensitivity maps for correctly classified cases in the three tasks. These maps highlight the regions of interest that contributed most to the model's predictions.

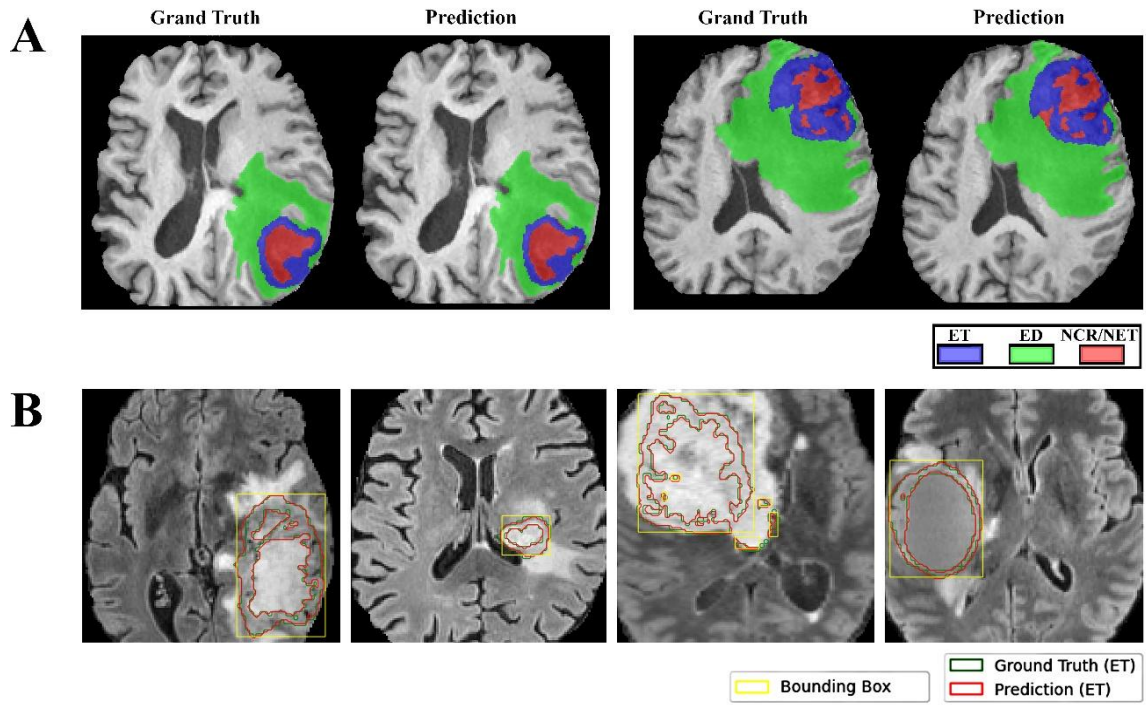

**Figure 3.** Segmentation results from two external test datasets. **(A)** Comparison of ground truth and predicted segmentation for different tumor subregions (ET, ED, NCR/NET) on FLAIR MRI axial slices from the UPenn-GBM dataset. **(B)** Predicted segmentation of the ET label with corresponding ground truth and bounding boxes on FLAIR MRI slices from the Ivy Gap dataset.

### 3.2.1. Ablation Experimental Results

To perform the ablation study of our proposed MTS-UNET network, we used the IDH mutation classification task, as it includes all four MRI modalities, and incorporates the CMD module. Additionally, we used the EGD test cohort, which offers a more balanced class distribution (IDH-mutant: 139, IDH-wild type: 276) compared to other test sets, except when specifically analyzing the impact of class distribution.

#### 3.2.1.1. Evaluation of Proposed Modules

Table 4 provides the ablation study results, evaluating the individual and combined effects of the TAFE and CMD modules on IDH prediction. For the TAFE module, we used the TAFE-2 set-up, which aggregates features from stages 3 and 4 of the encoder ($x_3$ and $x_4$). When used alone, TAFE achieved an AUC of 84.38%, while the CMD module performed slightly better across all metrics. Notably,

Table 3. The mean ± standard deviation of evaluation metrics for prediction models across 5-fold cross validation and external test cohorts: TCGA (The Cancer Genome Atlas), Ivy GAP (Ivy Glioblastoma Atlas Project), RHUH-GBM (Río Hortega University Hospital Glioblastoma Dataset), UPenn-GBM (University of Pennsylvania Glioblastoma Dataset), UCSF-PDGM (University of California San Francisco Preoperative Diffuse Glioma MRI), EGD (Erasmus Glioma Database), and LGG-1p19q (LGG-1p19q Deletion). For IDH prediction, TCGA and UCSF-PDGM were used for cross-validation. The UCSF-PDGM, Ivy GAP, and LGG-1p19q cohorts were used for training and internal validation of 1p/19q co-deletion and grading tasks, while RHUH-GBM and UPenn-GBM were used only for grading since they lacked 1p/19q co-deletion labels. The best-performing mean values are highlighted in **bold**. Asterisks indicate statistical significance compared to MTS-UNET, where *p < 0.05, **p < 0.001, and ***p < 0.0001.

**Internal Validation**

| Dataset | Model | IDH | | | | 1p/19q | | | | Grade | | | |
|---|---|---|---|---|---|---|---|---|---|---|---|---|---|
| | | ACC (mean±std) (%) | F1 (mean±std) (%) | MCC (mean±std) (%) | AUC (mean±std) (%) | ACC (mean±std) (%) | F1 (mean±std) (%) | MCC (mean±std) (%) | AUC (mean±std) (%) | ACC (mean±std) (%) | F1 (mean±std) (%) | MCC (mean±std) (%) | AUC (mean±std) (%) |
| Internal Validation | ResNet10 | 71.10 ± 2.19 | 59.77 ± 5.03 | 41.01 ± 4.04 | 82.31 ± 3.07* | 86.03 ± 3.73 | 70.08 ± 4.99 | 62.10 ± 3.05 | 87.00 ± 4.08 | 81.73 ± 3.16 | 87.16 ± 2.17 | 55.02 ± 2.13 | 76.10 ± 3.29* |
| | ResNet50 | 74.17 ± 11.14 | 67.53 ± 23.44 | 43.40 ± 10.33 | 78.60 ± 11.62 | 87.5 ± 3.02 | 70.33 ± 7.70 | 62.57 ± 5.21 | 87.60 ± 5.73 | 75.82 ± 6.12 | 84.18 ± 3.38 | 46.05 ± 4.12 | 70.07 ± 12.38 |
| | SENet101 | 74.43 ± 6.63 | 70.38 ± 14.71 | 47.64 ± 8.21 | 81.38 ± 3.64* | 84.75 ± 4.41 | 58.49 ± 28.50 | 52.72 ± 13.02 | 85.22 ± 4.54 | 80.57 ± 3.57 | 86.08 ± 2.52 | 55.90 ± 3.02 | 83.30 ± 2.72* |
| | DenseNet121 | 70.03 ± 4.71 | 62.79 ± 3.04 | 40.08 ± 3.98 | 73.65 ± 4.16* | 88.05 ± 1.72 | 72.36 ± 12.45 | 65.61 ± 5.04 | 89.36 ± 2.86 | 79.00 ± 3.58 | 83.62 ± 4.73 | 53.10 ± 4.12 | 85.14 ± 4.21* |
| | ViT-4 | 80.21 ± 5.03 | 81.19 ± 5.56 | 61.10 ± 6.02 | 88.83 ± 7.04* | 85.48 ± 4.81 | 70.40 ± 7.06 | 62.50 ± 3.64 | 86.75 ± 3.14* | 80.21 ± 4.13 | 86.27 ± 3.17 | 55.32 ± 5.07 | 80.01 ± 2.55* |
| | ViT-16 | 78.12 ± 6.19 | 76.34 ± 8.44 | 56.4 ± 6.12 | 85.83 ± 2.73 | **88.78 ± 0.40** | 74.46 ± 1.18 | 67.81 + 2.02 | 85.73 ± 1.63 | 81.87 ± 2.89 | 86.90 ± 2.22 | 59.13 ± 3.65 | 80.07 ± 5.46* |
| | MTS-UNET | **90.88 ± 2.35** | **90.89 ± 2.35** | **82.06 ± 4.67** | **93.31 ± 2.46** | 88.42 ± 4.03 | **74.33 ± 5.68** | **67.56 ± 7.49** | **92.17 ± 3.09** | **89.35 ± 5.83** | **91.86 ± 4.25** | **76.63 ± 13.39** | **95.70 ± 3.26** |

**External Validation**

| Dataset | Model | IDH | | | | 1p/19q | | | | Grade | | | |
|---|---|---|---|---|---|---|---|---|---|---|---|---|---|
| | | ACC | F1 | MCC | AUC | ACC | F1 | MCC | AUC | ACC | F1 | MCC | AUC |
| EGD | ResNet10 | 56.56 ± 1.65 | 52.20 ± 2.32 | 13.43 ± 3.36 | 56.42 ± 1.79*** | 55.26 ± 10.68 | 30.01 ± 21.62 | 4.028 ± 8.12 | 54.22 ± 5.93 | 48.61 ± 2.43 | 47.70 ± 3.08 | 4.57 ± 5.39 | 54.30 ± 4.75* |
| | ResNet50 | 55.01 ± 1.26 | 60.37 ± 3.96 | 10.78 ± 2.84 | 53.43 ± 2.75** | **64.25 ± 2.38** | 24.95 ± 11.52 | 5.48 ± 8.63 | 61.47 ± 6.04 | 60.57 ± 5.05 | 73.47 ± 8.55 | 5.08 ± 3.23 | 48.00 ± 0.64** |
| | SENet101 | 52.35 ± 4.66 | 49.10 ± 3.90 | 12.25 ± 8.10 | 60.31 ± 4.94** | 59.00 ± 5.60 | 38.01 ± 4.35 | 7.74 ± 6.45 | 53.63 ± 2.67** | 50.47 ± 2.55 | 46.46 ± 2.05 | 12.46 ± 8.15 | 56.76 ± 5.11* |
| | DenseNet121 | 62.64 ± 4.68 | 50.89 ± 6.91 | 22.21 ± 8.77 | 64.52 ± 4.58* | 61.61 ± 6.20 | 35.42 ± 9.84 | 10.35 ± 4.43 | 60.13 ± 3.33 | 55.19 ± 3.55 | 60.75 ± 9.42 | 8.53 ± 1.80 | 58.60 ± 2.80* |
| | ViT-4 | 75.01 ± 5.58 | 64.67 ± 7.04 | 46.69 ± 9.57 | 80.56 ± 5.76* | 51.11 ± 5.26 | 40.41 ± 3.86 | 3.42 ± 1.87 | 50.38 ± 2.61* | 61.00 ± 6.83 | 72.75 ± 9.40 | 4.35 ± 7.61 | 45.01 ± 1.12** |
| | ViT-16 | 72.90 ± 7.61 | 53.95 ± 15.12 | 37.45 ± 13.57 | 75.20 ± 7.60* | 45.05 ± 2.01 | 44.09 ± 3.31 | 2.53 ± 3.56 | 51.06 ± 1.88** | 46.45 ± 2.44 | 43.81 ± 8.95 | 0.88 ± 3.93 | 48.07 ± 2.69** |
| | MTS-UNET | **83.23 ± 1.27** | **83.70 ± 0.54** | **67.01 ± 2.10** | **90.58 ± 1.25** | 61.65 ± 8.01 | **56.79 ± 11.19** | **25.28 ± 4.86** | **69.22 ± 3.58** | **77.95 ± 4.76** | **84.27 ± 2.69** | **52.04 ± 10.83** | **87.54 ± 2.65** |
| TCGA | ResNet10 | - | - | - | - | 50.72 ± 26.64 | 15.22 ± 11.07 | 2.04 ± 8.74 | 50.36 ± 6.87 | 48.08 ± 0.81 | 14.60 ± 9.12 | 1.70 ± 5.97 | 44.46 ± 3.14* |
| | ResNet50 | - | - | - | - | **76.84 ± 6.95** | 3.00 ± 1.83 | 2.22 ± 7.96 | 46.64 ± 4.70* | 51.18 ± 1.02 | 57.86 ± 19.96 | 3.60 ± 3.01 | 53.86 ± 5.80 |
| | SENet101 | - | - | - | - | 50.00 ± 9.23 | 22.11 ± 16.61 | 3.56 ± 5.89 | 49.90 ± 6.20 | 52.33 ± 1.53 | 27.43 ± 5.52 | 7.24 ± 4.59 | 49.00 ± 2.17* |
| | DenseNet121 | - | - | - | - | 43.35 ± 15.13 | 21.30 ± 4.97 | 3.08 ± 3.52 | 53.86 ± 3.52 | 51.14 ± 0.54 | 11.61 ± 11.52 | 7.82 ± 3.00 | 48.28 ± 0.96* |
| | ViT-4 | - | - | - | - | 49.00 ± 12.53 | 22.00 ± 0.58 | 2.19 ± 1.37 | 50.21 ± 1.26* | 45.51 ± 1.28 | 22.53 ± 9.94 | 3.78 ± 11.1 | 40.11 ± 2.16* |
| | ViT-16 | - | - | - | - | 60.01 ± 24.32 | 18.87 ± 2.11 | 1.01 ± 3.68 | 51.41 ± 2.65 | 47.08 ± 0.98 | 38.31 ± 12.34 | 1.10 ± 8.11 | 44.24 ± 1.48* |
| | MTS-UNET | - | - | - | - | 71.19 ± 3.51 | **35.02 ± 5.76** | **24.03 ± 8.37** | **67.12 ± 6.40** | **72.09 ± 7.04** | **73.05 ± 5.78** | **46.02 ± 21.81** | **77.37 ± 7.38** |
| Ivy GAP, RHUH-GBM | ResNet10 | 52.02 ± 1.92 | 26.05 ± 3.48 | 6.55 ± 6.08 | 38.44 ± 3.33* | - | - | - | - | - | - | - | - |
| | ResNet50 | 56.09 ± 4.94 | 34.16 ± 3.12 | 18.44 ± 13.61 | 41.01 ± 21.27 | - | - | - | - | - | - | - | - |
| | SENet101 | 53.19 ± 2.15 | 24.48 ± 8.65 | 11.30 ± 7.29 | 49.46 ± 6.49* | - | - | - | - | - | - | - | - |
| | DenseNet121 | 51.30 ± 1.79 | 11.44 ± 8.54 | 9.62 ± 9.91 | 50.45 ± 1.75* | - | - | - | - | - | - | - | - |
| | ViT-4 | 54.78 ± 3.87 | 60.15 ± 10.92 | 10.61 ± 8.79 | 51.88 ± 2.41 | - | - | - | - | - | - | - | - |
| | ViT-16 | 60.72 ± 5.35 | 53.89 ± 18.22 | 20.83 ± 13.01 | 59.15 ± 3.95 | - | - | - | - | - | - | - | - |
| | MTS-UNET | **66.04 ± 2.49** | **61.09 ± 2.03** | **33.51 ± 5.56** | **65.41 ± 3.35** | - | - | - | - | - | - | - | - |
| UPenn-GBM | ResNet10 | **95.38 ± 1.74** | 6.82 ± 3.03 | 5.53 ± 2.81 | 52.69 ± 3.74 | - | - | - | - | - | - | - | - |
| | ResNet50 | 89.71 ± 3.14 | 7.41 ± 2.19 | 2.32 ± 2.38 | 37.82 ± 7.56* | - | - | - | - | - | - | - | - |
| | SENet101 | 81.40 ± 8.18 | 8.09 ± 9.06 | 6.22 ± 4.78 | 63.17 ± 8.88 | - | - | - | - | - | - | - | - |
| | DenseNet121 | 90.61 ± 3.11 | 4.92 ± 3.38 | 1.65 ± 1.87 | 55.07 ± 1.98 | - | - | - | - | - | - | - | - |
| | ViT-4 | 63.43 ± 11.14 | 11.22 ± 2.21 | 13.45 ± 4.66 | 75.17 ± 7.06 | - | - | - | - | - | - | - | - |
| | ViT-16 | 73.74 ± 16.03 | 11.68 ± 4.34 | 12.76 ± 6.11 | 60.66 ± 8.43 | - | - | - | - | - | - | - | - |
| | MTS-UNET | 86.81 ± 4.47 | **24.11 ± 4.44** | **26.30 ± 4.14** | **80.31 ± 1.09** | - | - | - | - | - | - | - | - |

**Abbreviations:** ACC, accuracy; AUC, the area under the curve; F1, F1-score; MCC, Matthews correlation coefficient.

integrating both modules through DSF further enhanced the model's performance, achieving an AUC of 90.58% and reducing variance across five runs. However, there was no statistically significant difference between the TAFE-only, CMD-only, and DSF methods.

Figure 5 shows Grad-CAM visualizations for an IDH mutant case with the T2-FLAIR mismatch sign. The T2-GCAM and FLAIR-GCAM panels highlight key regions influencing classification. Using CMD alone focuses on mismatch areas (e.g., T2 hyperintensities or FLAIR rim-like patterns), while DSF maps, which incorporate TAFE features, expand attention to a broader tumor region, enhancing both mismatch-related and tumor-focused representations.

**Table 4.** Ablation studies on the three proposed modules for IDH prediction on the EGD (Erasmus Glioma Database) test dataset. There is no statistically significant difference between methods.

| TAFE | CMD | ACC (mean±std) (%) | F1 (mean±std) (%) | MCC (mean±std) (%) | AUC (mean±std) (%) | Parameter Size |
|---|---|---|---|---|---|---|
| ✓ |   | 76.34 ± 8.59 | 74.53 ± 12.75 | 54.42 ± 15.70 | 84.38 ± 7.78 | 62 M |
|   | ✓ | 79.28 ± 4.92 | 78.50 ± 8.63 | 60.04 ± 8.08 | 85.97 ± 2.44 | 62 M |
| ✓ | ✓ | 83.23 ± 1.27 | 83.70 ± 0.54 | 67.01± 2.10 | 90.58 ± 1.25 | 124 M |

**Abbreviations**: ACC, accuracy; AUC, the area under the curve; F1, F1-score; MCC, Matthews correlation coefficient.

### 3.2.1.2. Segmentation Guidance and Feature Aggregation in TAFE

We assessed the impact of segmentation guidance and feature aggregation depth on the performance of the TAFE module by comparing two setups: the TAFE module, which incorporates segmentation guidance, and a baseline Swin Transformer (SwinT) without it (Table 5). Each module was tested under four conditions: using only the deepest encoder feature ($x_4$, labeled "1") and aggregating features from two ($x_3$ and $x_4$, labeled "2"), three ($x_2$ to $x_4$, labeled "3"), and all four encoder stages ($x_1$ to $x_4$, labeled "4") on the EGD dataset. The results show that segmentation guidance consistently enhances performance across all metrics. For instance, TAFE–1 achieved 74.19% accuracy and an AUC of 86.62%, significantly outperforming SwinT–1, which reached 70.98% accuracy and an AUC of 75.38% ($p < 0.05$). Moreover, feature aggregation adversely affected SwinT, reducing its accuracy to 53.30% and AUC to 61.86% ($p < 0.001$ vs. TAFE–1). As shown in figure 6, occlusion sensitivity maps for one IDH-mutant case reveal that both TAFE–1 and TAFE–4 accurately predicted IDH status by focusing on key tumor regions, whereas SwinT–4 misclassified the case by concentrating on non-tumoral areas. These findings collectively highlight the importance of segmentation-derived information in guiding feature extraction. Notably, varying the aggregation depth within the TAFE module did not yield significant performance differences.

### 3.2.1.3. MRI Sequence Combinations

Table 6 summarizes the performance of the TAFE and DSF modules across different MRI sequence combinations. While the TAFE module exhibited some variability in validation metrics depending on the input sequences, the DSF module consistently maintained high performance. The highest performance was observed when all four MRI modalities (T1, T1C, T2, and FLAIR) were included, though differences across sequence combinations were not statistically significant.

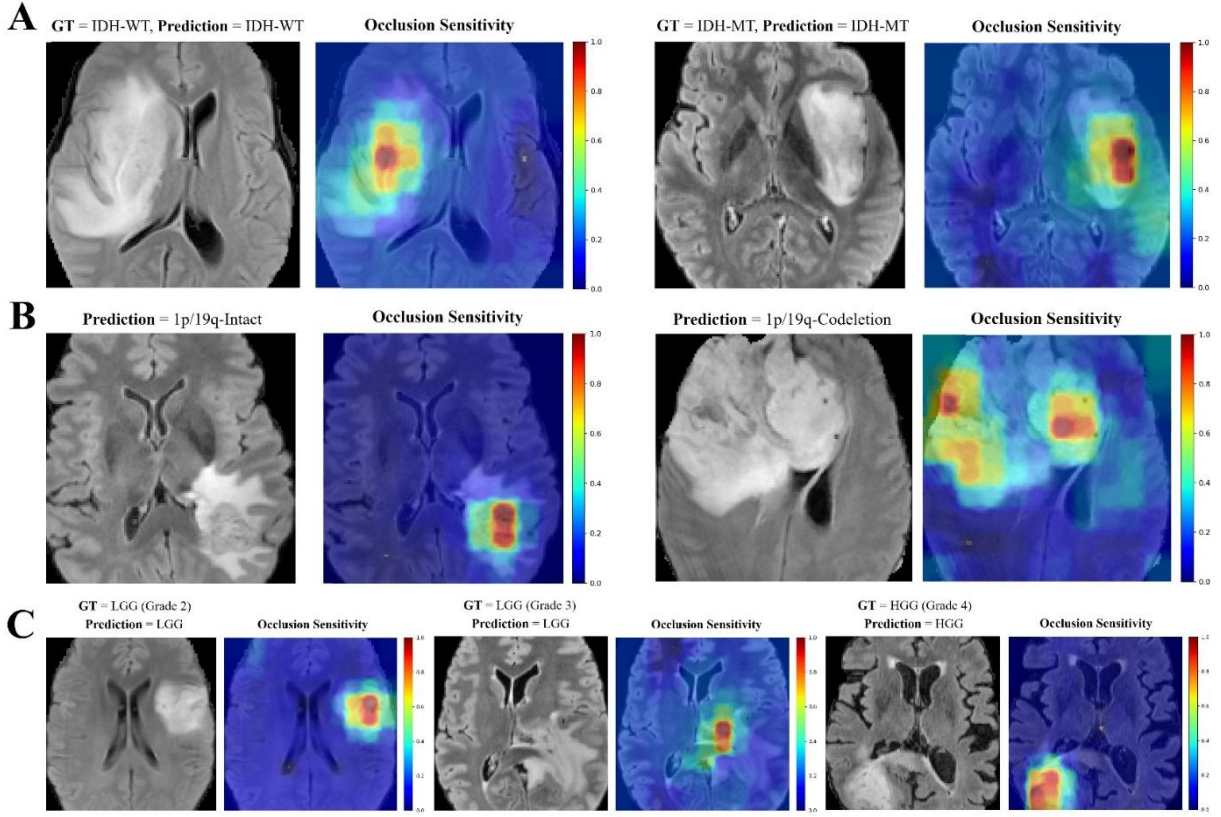

**Figure 4.** Occlusion sensitivity maps for correctly predicted **(A)** IDH mutation, **(B)** 1p/19q codeletion, and **(C)** tumor grade, overlaid on FLAIR MRI slices.

**Table 5.** Comparison of segmentation guidance and feature aggregation depth on the TAFE module performance for IDH mutation prediction on the EGD (Erasmus Glioma Database) test dataset. The SwinT refers to the encoder without segmentation guidance, and the TAFE refers to the encoder with segmentation guidance. TAFE–1 and SwinT–1 use the deepest encoder feature ($x_4$), while TAFE–4 and SwinT–4 aggregate features from all four stages ($x_1$ to $x_4$). The best-performing metrics (mean values) are highlighted in **bold**. '*' denotes pairwise statistical significance between TAFE and its corresponding SwinT module, while '†' indicates statistical significance compared to the SwinT-4 method. The number of symbols represents the significance level: *p < 0.05.

| Model Configuration | ACC (mean±std) (%) | F1 (mean±std) (%) | MCC (mean±std) (%) | AUC (mean±std) (%) |
|---|---|---|---|---|
| TAFE–1 | 74.19 ± 3.73 | 73.34 ± 7.82 | 50.86 ± 5.60 | **86.62 ± 1.67**† |
| SwinT-1 | 70.98 ± 2.82 | 60.74 ± 2.69 | 39.25 ± 4.00 | 75.38 ± 3.87*† |
| TAFE–2 | **76.34 ± 8.59** | **74.53 ± 12.75** | **54.42 ± 15.70** | 84.38 ± 7.78† |
| SwinT-2 | 65.64 ± 7.89 | 49.70 ± 10.89 | 24.95 ± 15.89 | 65.66 ± 10.45 |
| TAFE–3 | 75.05 ± 3.66 | 73.56 ± 9.73 | 51.08 ± 12.95 | 80.72 ± 7.42 |
| SwinT-3 | 65.98 ± 11.57 | 55.80 ± 8.91 | 30.01 ± 18.54 | 68.73 ± 12.78 |
| TAFE–4 | 73.69 ± 7.05 | 74.49 ± 5.42 | 48.23 ± 14.03 | 80.41 ± 7.58 |
| SwinT-4 | 53.30 ± 8.22 | 50.16 ± 4.67 | 16.50 ± 10.86 | 61.86 ± 6.03 |

**Abbreviations:** ACC, accuracy; AUC, the area under the curve; F1, F1-score; MCC, Matthews correlation coefficient.

### 3.2.1.4. Class distribution

Table 7 illustrates the impact of class skewness in test cohorts on the model performance. To optimize dataset utilization, two scenarios were designed: (1) using the TCGA and UCSF-PDGM cohorts for training and internal validation, with the remaining sets serving as independent test sets; and (2) using the EGD and UCSF-PDGM datasets for training and internal validation, with TCGA as an external validation set. The more balanced EGD and TCGA cohorts yielded stronger predictive metrics, with no significant difference between them, whereas the highly imbalanced Ivy GAP, RHUH-GBM, and UPenn-GBM cohorts demonstrated significantly lower performance.

**Table 6.** Impact of MRI sequence combinations as TAFE module inputs on the IDH prediction of the TAFE and DSF modules for the EGD (Erasmus Glioma Database) test cohort. There is no statistically significant difference between different sequence combination inputs.

| Sequence Combination | TAFE | | | | DSF | | | |
|---|---|---|---|---|---|---|---|---|
| | ACC (mean±std) (%) | F1 (mean±std) (%) | MCC (mean±std) (%) | AUC (mean±std) (%) | ACC (mean±std) (%) | F1 (mean±std) (%) | MCC (mean±std) (%) | AUC (mean±std) (%) |
| T1, T2 | 70.90 ± 12.05 | 75.87 ± 5.52 | 53.45 ± 5.19 | 78.04 ± 15.64 | 81.86 ± 3.73 | 82.04 ± 4.47 | 64.41 ± 7.26 | 89.25 ± 2.68 |
| T1C, T2 | 65.02 ± 4.05 | 71.83 ± 2.76 | 35.07 ± 7.65 | 79.80 ± 3.88 | 79.12 ± 2.16 | 80.22 ± 2.70 | 59.91 ± 3.23 | 88.32 ± 0.88 |
| T1C, FLAIR | **78.14 ± 3.62** | **78.04 ± 4.56** | **56.59 ± 7.43** | **85.01 ± 3.81** | **83.37 ± 2.04** | **84.01 ± 1.30** | 67.34 ± 3.37 | 90.51 ± 1.16 |
| T1, T1C, T2 | 67.05 ± 10.08 | 71.44 ± 5.59 | 35.47 ± 19.42 | 76.75 ± 6.77 | 79.06 ± 5.41 | 80.45 ± 3.12 | 59.76 ± 8.19 | 89.01 ± 1.63 |
| T1, T1C, FLAIR | 73.62 ± 6.45 | 76.24 ± 3.72 | 48.38 ± 11.38 | 83.41 ± 8.84 | 81.01 ± 4.42 | 80.93 ± 6.13 | 62.73 ± 8.55 | 88.90 ± 3.81 |
| T1, T1C, T2, FLAIR | 76.34 ± 8.59 | 74.53 ± 12.75 | 54.42 ± 15.70 | 84.38 ± 7.78 | 83.23 ± 1.27 | 83.70 ± 0.54 | **67.01± 2.10** | **90.58 ± 1.25** |

**Abbreviations**: ACC, accuracy; AUC, the area under the curve; F1, F1-score; MCC, Matthews correlation coefficient.

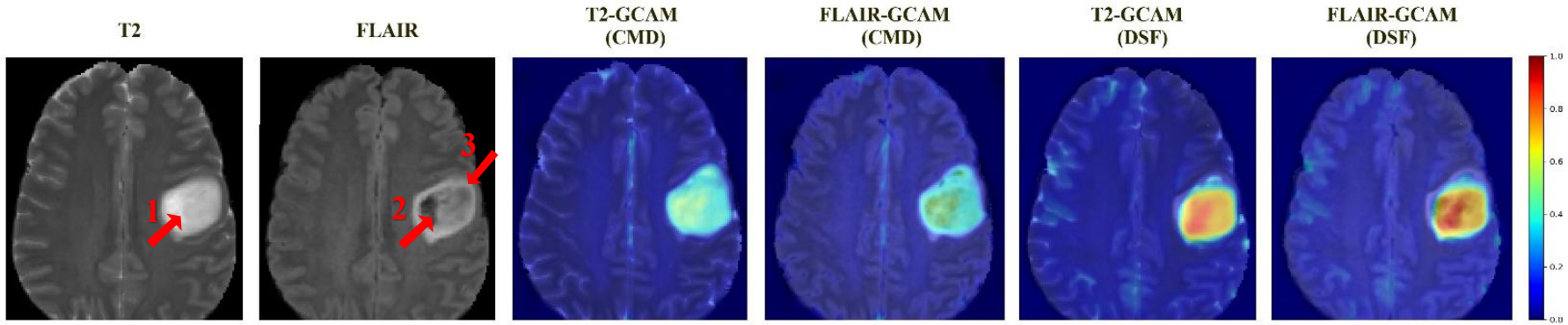

**Figure 5.** T2-FLAIR mismatch sign in one IDH mutant patient. Hyperintense signals on T2 are indicated by red arrows (1), while relatively hypointense signals and a hyperintense peripheral rim on FLAIR are marked by red arrows (2) and (3). To illustrate how the MTS-UNET network detects this sign, Grad-CAM maps (denoted as GCAM) are presented, where T2-GCAM represents the Grad-CAM map overlaid on the T2 scan, and FLAIR-GCAM represents the Grad-CAM map overlaid on the FLAIR scan. To enhance the interpretability of the two modules, Grad-CAMs obtained from the CMD module alone (T2-GCAM (CMD) and FLAIR-GCAM (CMD)) and from the fusion of both CMD and TAFE modules (T2-GCAM (DSF) and FLAIR-GCAM (DSF)) are provided.

**Table 7.** Comparison of dataset skewness (class distribution) on model performance. Two scenarios were designed to maximize dataset utilization: (1) using the TCGA (The Cancer Genome Atlas) and UCSF-PDGM (University of California San Francisco Preoperative Diffuse Glioma MRI) cohorts for training and internal validation (rows 1, 3, and 4); and (2) using the EGD (Erasmus Glioma Database) and UCSF datasets for training and internal validation, with TCGA as an independent test set (row 2). Class distributions for IDH status are presented as counts and percentages. Best-performing metrics (mean values) are highlighted in **bold**. '*' denotes significance compared to the EGD dataset, while '†' denotes significance compared to the TCGA cohort. The number of symbols represents the significance level: *p < 0.05 and **p < 0.001.

| Test Set | IDH Status (Class distribution) | ACC (mean±std) (%) | F1 (mean±std) (%) | MCC (mean±std) (%) | AUC (mean±std) (%) |
|---|---|---|---|---|---|
| EGD | MT: 139 (33%) WT: 276 (67%) | 83.23 ± 1.27 | **83.70 ± 0.54** | **67.01 ± 2.10** | **90.58 ± 1.25** |
| TCGA | MT: 89 (42%) WT: 124 (58%) | 81.22 ± 3.60 | 79.17 ± 3.34 | 62.79 ± 6.56 | 88.08 ± 3.08 |
| Ivy GAP, RHUH-GBM | MT: 7 (9%) WT: 67 (91%) | 66.04 ± 2.49 | 61.09 ± 2.03 | 33.51 ± 5.56 | 65.41 ± 3.35** †† |
| UPenn-GBM | MT: 16 (3%) WT: 498 (97%) | **86.81 ± 4.47** | 24.11 ± 4.44 | 26.30 ± 4.14 | 80.31 ± 1.09* † |

**Abbreviations:** ACC, accuracy; AUC, the area under the curve; F1, F1-score; MCC, Matthews correlation coefficient; IDH, isocitrate dehydrogenase; MT, IDH-mutated; WT, IDH-wildtype.

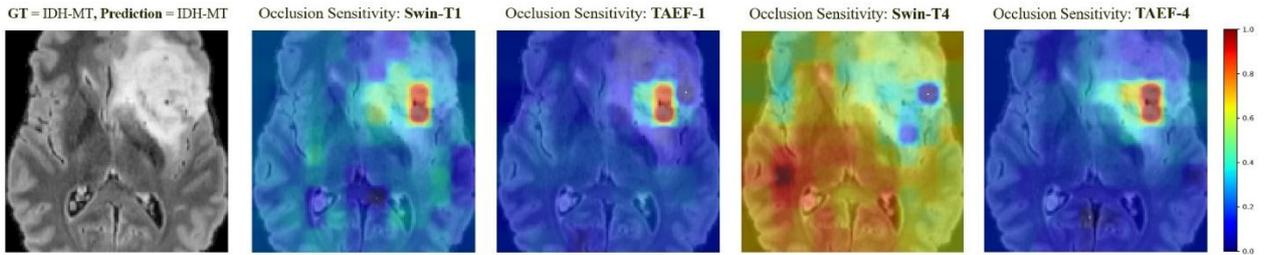

**Figure 6.** Comparison of occlusion sensitivity maps for the multi-level, multi-scale TAEF module with segmentation guidance (TAFE) and without it (Swin-T) under two specific setups: Swin-T1, TAEF-1, Swin-T4, and TAEF-4. Specifically, Swin-T1 and TAEF-1 refer to models that use only the deepest encoder feature ($x_4$), while Swin-T4 and TAEF-4 aggregate features from all four encoder stages ($x_1$ to $x_4$). The heatmaps, overlaid on a FLAIR MRI slice, highlight key regions used for IDH prediction. For the same patient, Swin-T1, TAEF-1, and TAEF-4 correctly predicted the IDH status, whereas Swin-T4 misclassified it. The heatmap for Swin-T4 suggests that it focused on non-tumoral regions, leading to the incorrect prediction. This underscores the importance of segmentation guidance and feature aggregation in improving IDH classification performance.

## 4. Discussion

In this study, we introduce a novel multi-task framework, MTS-UNET, which integrates glioma segmentation, molecular subtyping (IDH mutation and 1p/19q co-deletion), and histological grading into a single, end-to-end deep learning model. Our method incorporates two key modules, TAFE and CMD, to extract both tumor-specific features and subtle inter-modality cues, addressing longstanding challenges in noninvasive glioma evaluation. By utilizing a foundation model pretrained on large-scale neuroimaging datasets, our approach effectively mitigates issues related to intratumoral heterogeneity and variability across multi-center MRI acquisitions.

The segmentation branch of our framework is based on the SWIN-UNETR architecture, built on the robust BrainSegFounder model [26]. This foundation model was pretrained using a two-stage strategy on large-scale, multi-modal MRI datasets, making it highly optimized for brain tumor segmentation without requiring additional architectural modifications. For fine-tuning on glioma segmentation, the model achieved a mean Dice score of 87% ± 1.82 in 5-fold cross-validation on the UCSF-PDGM dataset. External validation on highly heterogeneous multi-center cohorts (Ivy GAP, UPenn-GBM, and RHUH-GBM) confirmed its robustness, with a mean Dice score of 84% ± 3.51. These results, along with a low Hausdorff distance and a high IoU, demonstrate that the network can accurately delineate tumor boundaries and capture fine-grained subregional details, such as ET, ED, and NCR/NET, even in the face of significant imaging variability. These results match or, in some cases, exceed those reported by the state-of-the-art transformer-based models [40] and muti-task segmentation networks [17,42,45]. The high segmentation accuracy is essential, as it forms the basis for the subsequent extraction of tumor-specific features and enhances the reliability of downstream molecular and histological predictions.

The classification results further validate the efficacy of our multi-task framework. MTS-UNET outperformed multiple baseline models, including ViT variants—the only specialized foundation-based model for predicting IDH mutation, 1p/19q co-deletion, and detecting brain tumors [25]. While some models achieved competitive results during cross-validation, their performance dropped significantly on external test sets, especially in balanced metrics like MCC and F1-score. This decline likely reflects their limited capacity to generalize to data with different imaging characteristics—a limitation that our segmentation-guided approach appears to overcome. These findings align with recent multi-task studies [17,18,42,46], which emphasize the benefits of simultaneously analyzing tumor localization and genotype. However, our study further demonstrates that a foundation model-based approach enhances both stability and generalizability across diverse patient cohorts.

Moreover, our model achieves higher performance in predicting IDH mutation compared to 1p/19q co-deletion and histological grading. Two factors seem to contribute to this difference. First, IDH prediction benefits from the combined use of both the TAFE and CMD modules. Since the T2–FLAIR mismatch sign is clinically associated with IDH status, the complementary cues provided by the CMD module enable MTS-UNET to more effectively distinguish between IDH-mutant and IDH-wildtype tumors. In contrast, the tasks of predicting 1p/19q co-deletion and performing histological grading rely solely on the TAFE module. Second, the IDH task utilized all four MRI modalities in both training and test datasets which provided the network with richer image features. Conversely, for 1p/19q co-deletion and grading, only T1C and T2 sequences were available, as the LGG–1p/19q dataset, which contained the most minority-class samples, did not include all four modalities. This narrower set of imaging inputs likely constrained the model's ability to capture subtle features tied to 1p/19q co-deletion and tumor grade.

Although MTS-UNET consistently outperforms other architectures, it still faces challenges with highly imbalanced test sets. For example, in IDH prediction, some external datasets exhibit extreme class imbalances—such as Ivy GAP, RHUH-GBM (91% IDH mutant vs. 9% IDH wildtype), and UPenn-GBM (only 16 IDH-mutant cases versus 498 IDH-wildtype). These extreme distributions result in significantly lower performance compared to the more balanced EGD and TCGA cohorts, which, although still imbalanced, are less skewed. Nevertheless, our network generally maintains significantly better predictive power than baseline approaches across all datasets. The challenge is even greater for 1p/19q co-deletion, where the scarcity of co-deleted cases further strains the model's ability to generalize. Despite these challenges, our tumor-centric strategy provides a robust framework that appears less sensitive to data shifts and imbalances than conventional classification-only models. Occlusion sensitivity maps further confirm that the model consistently focuses on clinically relevant regions, adding an extra layer of confidence in its predictions.

Our ablation experiments provide important insights into the contributions of the individual modules within the MTS-UNET framework. When evaluated on the IDH mutation prediction task, the TAFE module yielded an AUC of 84.38% when used in isolation, while the CMD module achieved slightly better performance. However, combining both modules led to a higher AUC of 90.58% and reduced variability across runs, in line with previous study [19]. Although the differences between individual modules and their fused approach were not statistically significant, the trend suggests that integrating segmentation-derived features with cross-modality differentials enhances predictive performance. Grad-CAM visualizations further support this observation, showing that DSF-derived attention maps not only focus on key mismatch regions but also extend across a broader tumor context, which may enhance the overall discriminative power of the model.

Further investigations into the TAFE module confirmed the critical role of segmentation guidance. Compared to a baseline Swin Transformer model that lacked segmentation-derived inputs, TAFE-based configurations—especially when using the deepest encoder feature ($x_4$)—outperformed across all metrics. We also performed a series of experiments focusing on different feature aggregation depths. The results showed the precise depth of aggregation did not yield significant differences within the TAFE module. In contrast, the performance of the baseline model degraded as more features were aggregated, suggesting that without segmentation cues, additional features may introduce noise rather than useful features. This observation emphasizes the value of integrating anatomical localization directly into the feature extraction process—a strategy that aligns with recent research for more context-aware deep learning models [16, 19, 46].

Our experiments also assessed the model's robustness with respect to variations in available MRI sequences. Consistent with a published study [19], the TAFE module showed some sensitivity to input modalities, with the lowest accuracy and MCC observed for the T1C-T2 input pair. Interestingly, these modalities are the same MRI sequences used for 1p/19q co-deletion and grading tasks, which can further explain the model's inferior performance in these tasks compared to IDH status prediction. In contrast, the DSF module maintained consistently high performance across different input variations. This stability is due to the integration of the CMD component within this network, which ensures that T2 and FLAIR mismatch features are captured even when these modalities are not explicitly included as inputs to the TAFE. Notably, while the best performance was achieved using all four MRI modalities, the differences did not reach statistical significance.

Despite these promising results, some limitations remain. A key challenge is the model's reliance on accurate tumor segmentation—especially for the CMD module—to guide feature extraction, which may limit its applicability when segmentation quality is compromised. Future work could explore segmentation-independent approaches or enhanced self-supervised strategies to reduce this dependency. Additionally, the CMD module's sensitivity to the T2-FLAIR mismatch sign, while beneficial for detection, may be affected by subtle variations across patients and imaging protocols. Furthermore, although our model generalizes well across multiple datasets, its performance declines in highly skewed cohorts, indicating the need for further improvements in discriminatory power. While our framework currently focuses on key biomarkers such as IDH mutation and 1p/19q co-deletion, future extensions should include additional molecular markers (e.g., MGMT promoter methylation, ATRX status) to enhance its clinical utility. Finally, although the computational requirements were manageable on high-end hardware, additional optimization is necessary for real-time clinical deployment. Moving forward, we plan to integrate additional imaging modalities, such as diffusion-weighted and perfusion imaging, and incorporate multi-omics data to develop a more comprehensive framework for personalized glioma management.

## 5. Conclusion

In summary, our MTS-UNET framework demonstrates that a multimodal, foundation model-based approach can successfully consolidate glioma segmentation, molecular subtyping, and histological grading into a unified architecture. The integration of segmentation-guided feature encoding with cross-modality differential analysis enhances both predictive performance and the interpretability of the model's decision-making process. This novel approach effectively captures global tumor morphology as well as subtle imaging biomarkers, as confirmed by our large multi-center external validations.

Nonetheless, further refinement and prospective validation are needed to translate these promising findings into routine clinical practice.

**Funding:**

This research did not receive any specific grant from funding agencies in the public, commercial, or not-for-profit sectors.

**Conflict of interest:**

The authors declare that they have no conflict of interest.

**Data Availability:**

The datasets are available at the following links:

**TCGA-LGG** [27]: https://www.cancerimagingarchive.net/collection/tcga-lgg/
**TCGA-GBM** [27]: https://www.cancerimagingarchive.net/collection/tcga-gbm/
**UCSF-PDGM** [31]: https://www.cancerimagingarchive.net/collection/ucsf-pdgm/
**EGD** [32]: https://xnat.bmia.nl/REST/projects/egd/
**Ivy GAP** [28]: https://www.cancerimagingarchive.net/collection/ivygap/
**RHUH-GBM** [29]: https://www.cancerimagingarchive.net/collection/rhuh-gbm/
**UPenn-GBM** [30]: https://www.cancerimagingarchive.net/collection/upenn-gbm/
**LGG-1p19q** [33,34]: https://www.cancerimagingarchive.net/collection/lgg-1p19qdeletion/

**References:**

1. Mohan, G.; Subashini, M.M. MRI Based Medical Image Analysis: Survey on Brain Tumor Grade Classification. Biomedical Signal Processing and Control 2018, 39, 139–161, doi:10.1016/j.bspc.2017.07.007.

2. Louis, D.N.; Perry, A.; Wesseling, P.; Brat, D.J.; Cree, I.A.; Figarella-Branger, D.; Hawkins, C.; Ng, H.K.; Pfister, S.M.; Reifenberger, G.; et al. The 2021 WHO Classification of Tumors of the Central Nervous System: A Summary. Neuro-Oncology 2021, 23, 1231–1251, doi:10.1093/neuonc/noab106.

3. IDH Mutations Predict Longer Survival and Response to Temozolomide in Secondary Glioblastoma - SongTao - 2012 - Cancer Science - Wiley Online Library Available online: https://onlinelibrary.wiley.com/doi/full/10.1111/j.1349-7006.2011.02134.x (accessed on 7 February 2025).

4. Han, S.; Liu, Y.; Cai, S.J.; Qian, M.; Ding, J.; Larion, M.; Gilbert, M.R.; Yang, C. IDH Mutation in Glioma: Molecular Mechanisms and Potential Therapeutic Targets. Br J Cancer 2020, 122, 1580–1589, doi:10.1038/s41416-020-0814-x.

5. Guarnera, A.; Romano, A.; Moltoni, G.; Ius, T.; Palizzi, S.; Romano, A.; Bagatto, D.; Minniti, G.; Bozzao, A. The Role of Advanced MRI Sequences in the Diagnosis and Follow-Up of Adult Brainstem Gliomas: A Neuroradiological Review. Tomography 2023, 9, 1526–1537, doi:10.3390/tomography9040122.

6. Maynard, J.; Okuchi, S.; Wastling, S.; Busaidi, A.A.; Almossawi, O.; Mbatha, W.; Brandner, S.; Jaunmuktane, Z.; Koc, A.M.; Mancini, L.; et al. World Health Organization Grade II/III Glioma Molecular Status: Prediction by MRI Morphologic Features and Apparent Diffusion Coefficient. Radiology 2020, 296, 111–121, doi:10.1148/radiol.2020191832.

7. Park, J.E.; Kim, H.S.; Kim, N.; Park, S.Y.; Kim, Y.-H.; Kim, J.H. Spatiotemporal Heterogeneity in Multiparametric Physiologic MRI Is Associated with Patient Outcomes in IDH-Wildtype Glioblastoma. Clinical Cancer Research 2021, 27, 237–245, doi:10.1158/1078-0432.CCR-20-2156.

8. Tian, Q.; Yan, L.-F.; Zhang, X.; Zhang, X.; Hu, Y.-C.; Han, Y.; Liu, Z.-C.; Nan, H.-Y.; Sun, Q.; Sun, Y.-Z.; et al. Radiomics Strategy for Glioma Grading Using Texture Features from Multiparametric MRI. Journal of Magnetic Resonance Imaging 2018, 48, 1518–1528, doi:10.1002/jmri.26010.

9. Two-Stage Training Framework Using Multicontrast MRI Radiomics for IDH Mutation Status Prediction in Glioma | Radiology: Artificial Intelligence Available online: https://pubs.rsna.org/doi/full/10.1148/ryai.230218 (accessed on 14 February 2025).

10. Hosny, A.; Aerts, H.J.; Mak, R.H. Handcrafted versus Deep Learning Radiomics for Prediction of Cancer Therapy Response. The Lancet Digital Health 2019, 1, e106–e107, doi:10.1016/S2589-7500(19)30062-7.

11. Ge, C.; Gu, I.Y.-H.; Jakola, A.S.; Yang, J. Enlarged Training Dataset by Pairwise GANs for Molecular-Based Brain Tumor Classification. IEEE Access 2020, 8, 22560–22570, doi:10.1109/ACCESS.2020.2969805.

12. Li, Z.; Wang, Y.; Yu, J.; Guo, Y.; Cao, W. Deep Learning Based Radiomics (DLR) and Its Usage in Noninvasive IDH1 Prediction for Low Grade Glioma. Sci Rep 2017, 7, 5467, doi:10.1038/s41598-017-05848-2.

13. Chang, P.; Grinband, J.; Weinberg, B.D.; Bardis, M.; Khy, M.; Cadena, G.; Su, M.-Y.; Cha, S.; Filippi, C.G.; Bota, D.; et al. Deep-Learning Convolutional Neural Networks Accurately Classify Genetic Mutations in Gliomas. Am. J. Neuroradiol. 2018, 39, 1201–1207, doi:10.3174/ajnr.A5667.

14. Nalawade, S.; Murugesan, G.K.; Vejdani-Jahromi, M.; Fisicaro, R.A.; Yogananda, C.G.B.; Wagner, B.; Mickey, B.; Maher, E.; Pinho, M.C.; Fei, B.; et al. Classification of Brain Tumor Isocitrate Dehydrogenase Status Using MRI and Deep Learning. JMI 2019, 6, 046003, doi:10.1117/1.JMI.6.4.046003.

15. Farahani, S.; Hejazi, M.; Tabassum, M.; Ieva, A.D.; Mahdavifar, N.; Liu, S. Diagnostic Performance of Deep Learning for Predicting Gliomas' IDH and 1p/19q Status in MRI: A Systematic Review and Meta-Analysis 2024.

16. Cheng, J.; Liu, J.; Kuang, H.; Wang, J. A Fully Automated Multimodal MRI-Based Multi-Task Learning for Glioma Segmentation and IDH Genotyping. IEEE Trans Med Imaging 2022, 41, 1520–1532, doi:10.1109/TMI.2022.3142321.

17. Decuyper, M.; Bonte, S.; Deblaere, K.; Van Holen, R. Automated MRI Based Pipeline for Segmentation and Prediction of Grade, IDH Mutation and 1p19q Co-Deletion in Glioma. Comput Med Imaging Graph 2021, 88, 101831, doi:10.1016/j.compmedimag.2020.101831.

18. van der Voort, S.R.; Incekara, F.; Wijnenga, M.M.J.; Kapsas, G.; Gahrmann, R.; Schouten, J.W.; Nandoe Tewarie, R.; Lycklama, G.J.; De Witt Hamer, P.C.; Eijgelaar, R.S.; et al. Combined Molecular Subtyping, Grading, and Segmentation of Glioma Using Multi-Task Deep Learning. Neuro Oncol 2023, 25, 279–289, doi:10.1093/neuonc/noac166.

19. Zhang, J.; Cao, J.; Tang, F.; Xie, T.; Feng, Q.; Huang, M. Multi-Level Feature Exploration and Fusion Network for Prediction of IDH Status in Gliomas From MRI. IEEE J. Biomed. Health Inform. 2024, 28, 42–53, doi:10.1109/JBHI.2023.3279433.

20. Khan, W.; Leem, S.; See, K.B.; Wong, J.K.; Zhang, S.; Fang, R. A Comprehensive Survey of Foundation Models in Medicine. IEEE Reviews in Biomedical Engineering 2025, 1–20, doi:10.1109/RBME.2025.3531360.

21. A Study of Generative Large Language Model for Medical Research and Healthcare | Npj Digital Medicine Available online: https://www.nature.com/articles/s41746-023-00958-w (accessed on 14 February 2025).

22. Segment Anything in Medical Images | Nature Communications Available online: https://www.nature.com/articles/s41467-024-44824-z.

23. Zhang, S.; Xu, Y.; Usuyama, N.; Xu, H.; Bagga, J.; Tinn, R.; Preston, S.; Rao, R.; Wei, M.; Valluri, N.; et al. BiomedCLIP: A Multimodal Biomedical Foundation Model Pretrained from Fifteen Million Scientific Image-Text Pairs 2025.

24. Self-Supervised Learning for Medical Image Classification: A Systematic Review and Implementation Guidelines | Npj Digital Medicine Available online: https://www.nature.com/articles/s41746-023-00811-0 (accessed on 14 February 2025).

25. Chen, M.; Zhang, M.; Yin, L.; Ma, L.; Ding, R.; Zheng, T.; Yue, Q.; Lui, S.; Sun, H. Medical Image Foundation Models in Assisting Diagnosis of Brain Tumors: A Pilot Study. Eur Radiol 2024, 34, 6667–6679, doi:10.1007/s00330-024-10728-1.

26. Cox, J.; Liu, P.; Stolte, S.E.; Yang, Y.; Liu, K.; See, K.B.; Ju, H.; Fang, R. BrainSegFounder: Towards 3D Foundation Models for Neuroimage Segmentation. Medical Image Analysis 2024, 97, 103301, doi:10.1016/j.media.2024.103301.

27. Advancing The Cancer Genome Atlas Glioma MRI Collections with Expert Segmentation Labels and Radiomic Features | Scientific Data Available online: https://www.nature.com/articles/sdata2017117.

28. An Anatomic Transcriptional Atlas of Human Glioblastoma | Science Available online: https://www.science.org/doi/full/10.1126/science.aaf2666 (accessed on 8 February 2025).

29. Cepeda, S.; García-García, S.; Arrese, I.; Herrero, F.; Escudero, T.; Zamora, T.; Sarabia, R. The Río Hortega University Hospital Glioblastoma Dataset: A Comprehensive Collection of Preoperative, Early Postoperative and Recurrence MRI Scans (RHUH-GBM). Data in Brief 2023, 50, 109617, doi:10.1016/j.dib.2023.109617.

30. The University of Pennsylvania Glioblastoma (UPenn-GBM) Cohort: Advanced MRI, Clinical, Genomics, & Radiomics | Scientific Data Available online: https://www.nature.com/articles/s41597-022-01560-7.

31. Calabrese, E.; Villanueva-Meyer, J.E.; Rudie, J.D.; Rauschecker, A.M.; Baid, U.; Bakas, S.; Cha, S.; Mongan, J.T.; Hess, C.P. The University of California San Francisco Preoperative Diffuse Glioma MRI Dataset. Radiology: Artificial Intelligence 2022, 4, e220058, doi:10.1148/ryai.220058.

32. van der Voort, S.R.; Incekara, F.; Wijnenga, M.M.J.; Kapsas, G.; Gahrmann, R.; Schouten, J.W.; Dubbink, H.J.; Vincent, A.J.P.E.; van den Bent, M.J.; French, P.J.; et al. The Erasmus Glioma Database (EGD): Structural MRI Scans, WHO 2016 Subtypes, and Segmentations of 774 Patients with Glioma. Data in Brief 2021, 37, 107191, doi:10.1016/j.dib.2021.107191.

33. Akkus, Z.; Ali, I.; Sedlář, J.; Agrawal, J.P.; Parney, I.F.; Giannini, C.; Erickson, B.J. Predicting Deletion of Chromosomal Arms 1p/19q in Low-Grade Gliomas from MR Images Using Machine Intelligence. J Digit Imaging 2017, 30, 469–476, doi:10.1007/s10278-017-9984-3.

34. Clark, K.; Vendt, B.; Smith, K.; Freymann, J.; Kirby, J.; Koppel, P.; Moore, S.; Phillips, S.; Maffitt, D.; Pringle, M.; et al. The Cancer Imaging Archive (TCIA): Maintaining and Operating a Public Information Repository. J Digit Imaging 2013, 26, 1045–1057, doi:10.1007/s10278-013-9622-7.

35. Integrative Imaging Informatics for Cancer Research: Workflow Automation for Neuro-Oncology (I3CR-WANO) | JCO Clinical Cancer Informatics Available online: https://ascopubs.org/doi/full/10.1200/CCI.22.00177.

36. The UK Biobank Imaging Enhancement of 100,000 Participants: Rationale, Data Collection, Management and Future Directions | Nature Communications Available online: https://www.nature.com/articles/s41467-020-15948-9 (accessed on 8 February 2025).

37. Baid, U.; Ghodasara, S.; Mohan, S.; Bilello, M.; Calabrese, E.; Colak, E.; Farahani, K.; Kalpathy-Cramer, J.; Kitamura, F.C.; Pati, S.; et al. The RSNA-ASNR-MICCAI BraTS 2021 Benchmark on Brain Tumor Segmentation and Radiogenomic Classification 2021.

38. Jain, R.; Johnson, D.R.; Patel, S.H.; Castillo, M.; Smits, M.; van den Bent, M.J.; Chi, A.S.; Cahill, D.P. "Real World" Use of a Highly Reliable Imaging Sign: "T2-FLAIR Mismatch" for Identification of IDH Mutant Astrocytomas. Neuro-Oncology 2020, 22, 936–943, doi:10.1093/neuonc/noaa041.

39. DeLong, E.R.; DeLong, D.M.; Clarke-Pearson, D.L. Comparing the Areas under Two or More Correlated Receiver Operating Characteristic Curves: A Nonparametric Approach. Biometrics 1988, 44, 837–845, doi:10.2307/2531595.

40. Hatamizadeh, A.; Nath, V.; Tang, Y.; Yang, D.; Roth, H.R.; Xu, D. Swin UNETR: Swin Transformers for Semantic Segmentation of Brain Tumors in MRI Images. In Proceedings of the Brainlesion: Glioma, Multiple Sclerosis, Stroke and Traumatic Brain Injuries; Crimi, A., Bakas, S., Eds.; Springer International Publishing: Cham, 2022; pp. 272–284.

41. Hu, J.; Shen, L.; Sun, G. Squeeze-and-Excitation Networks.; 2018; pp. 7132–7141.

42. Biologically Interpretable Multi-Task Deep Learning Pipeline Predicts Molecular Alterations, Grade, and Prognosis in Glioma Patients | Npj Precision Oncology Available online: https://www.nature.com/articles/s41698-024-00670-2 (accessed on 8 February 2025).

43. Chang, K.; Bai, H.X.; Zhou, H.; Su, C.; Bi, W.L.; Agbodza, E.; Kavouridis, V.K.; Senders, J.T.; Boaro, A.; Beers, A.; et al. Residual Convolutional Neural Network for the Determination of IDH Status in Low- and High-Grade Gliomas from MR Imaging. Clinical Cancer Research 2018, 24, 1073–1081, doi:10.1158/1078-0432.CCR-17-2236.

44. Liang, S.; Zhang, R.; Liang, D.; Song, T.; Ai, T.; Xia, C.; Xia, L.; Wang, Y. Multimodal 3D DenseNet for IDH Genotype Prediction in Gliomas. Genes 2018, 9, 382, doi:10.3390/genes9080382.

45. Dorent, R.; Joutard, S.; Modat, M.; Ourselin, S.; Vercauteren, T. Hetero-Modal Variational Encoder-Decoder for Joint Modality Completion and Segmentation. In Proceedings of the Medical Image Computing and Computer Assisted Intervention – MICCAI 2019; Shen, D., Liu, T., Peters, T.M., Staib, L.H., Essert, C., Zhou, S., Yap, P.-T., Khan, A., Eds.; Springer International Publishing: Cham, 2019; pp. 74–82.

46. Chen, Q.; Wang, L.; Deng, Z.; Wang, R.; Wang, L.; Jian, C.; Zhu, Y.-M. Cooperative Multi-Task Learning and Interpretable Image Biomarkers for Glioma Grading and Molecular Subtyping. Medical Image Analysis 2025, 101, 103435, doi:10.1016/j.media.2024.103435.